\documentclass[a4paper,11pt]{article}

\usepackage{natbib}
\usepackage[colorlinks,bookmarksopen,bookmarksnumbered,citecolor=red,urlcolor=red]{hyperref}
\usepackage{authblk}
\usepackage{amsmath}
\usepackage{graphicx}

\newcommand\BibTeX{{\rmfamily B\kern-.05em \textsc{i\kern-.025em b}\kern-.08em
T\kern-.1667em\lower.7ex\hbox{E}\kern-.125emX}}

\usepackage{moreverb}

\begin{document}


\title{Quantile forecast discrimination ability and value}

\author[1,2]{Zied Ben Bouall\`egue}
\author[3]{Pierre Pinson}
\author[2]{Petra Friederichs}

\affil[1] {Deutscher Wetterdienst, Offenbach,Germany}
\affil[2] {Meteorological Institute, University of Bonn, Germany}
\affil[3] {Technical University of Denmark, Denmark}

\maketitle

\begin{abstract}
While probabilistic forecast verification for categorical forecasts is well established, some of the existing concepts and methods have not found their equivalent for the case of continuous variables. New tools dedicated to the assessment of forecast discrimination ability and forecast value are introduced here, based on quantile forecasts being the base product for the continuous case (hence in a nonparametric framework). The relative user characteristic (RUC) curve and the quantile value plot allow analysing the performance of a forecast for a specific user in a decision-making framework. The RUC curve is designed as a user-based discrimination tool and the quantile value plot translates forecast discrimination ability in terms of economic value. The relationship between the overall value of a quantile forecast and the respective quantile skill score is also discussed. The application of these new verification approaches and tools is illustrated based on synthetic datasets, as well as for the case of global radiation forecasts from the high resolution ensemble COSMO-DE-EPS of the German Weather Service.
\end{abstract}

\section{Introduction}
Verification of probabilistic weather forecasts is an area of intensive research and growing interest as ensemble forecasting is becoming a standard approach in numerical weather prediction. Ensemble prediction systems (EPS) issue a sample of possible future states of the atmosphere \citep{lewis05,leut08}. The forecasts can be interpreted in the form of a predictive distribution and probabilistic products can be derived in order to support and optimize forecast-based  decision-making \citep{krz83}. Appropriate tools for the assessment of probabilistic products from this perspective are therefore essential. 

Such tools already exist for probabilistic products expressed in the form a probability forecast.  The relative operating characteristic (ROC) curve  is a common verification tool for the assessment of probability forecasts \citep{mason82}. The ROC  curve is related to decision-making analysis and the corresponding  fundamental property of the forecast is called \textit{discrimination}. 
 Forecast discrimination assesses whether the forecast can be used to successfully discriminate between the observations \citep{murphy91} or, said differently, whether appropriate decisions can be taken based on a forecast.
Discrimination is translated in terms of \textit{economic value} using a simple cost-loss model that allows the specificity of a user to be taken into account through the definition of a \textit{cost-loss ratio}. The derived quantitative measure is called \textit{value score} or \textit{relative value} and is usually represented in the form of a probability value plot showing the forecast value as a function of the user's cost-loss ratio \citep{richardson2000,wilks2001,zhu02}. The value of a forecast is defined as the benefit to a user as a result of making decisions based on a forecast and has to be distinguished from forecast quality, the overall agreement between forecast and observation \citep{murphy93}. In a verification process, value and quality can be seen as being from the point of view of the forecast user and from the point of view of the forecast provider, respectively. The distinction between the two \textit{types of goodness}, value and quality, is crucial since a non-linear relationship between them can lead to situations where a large improvement in the forecast quality does not imply an increase in the forecast value, or conversely, a small improvement in  forecast quality can bring a notable benefit in terms of forecast value \citep{chen87,buizza2000,pinson2013}. 

Probabilistic products can be expressed in terms of a probability when the focus is on a particular event of interest, but also in terms of a quantile when the focus is on a particular probability level of interest.  While a probability forecast first requires the definition of an event, i.e. the categorization  of the original information, a quantile forecast is a 'single-valued' forecast expressed in the unit of the variable being forecast. Considering here probabilistic products derived from EPS simulations for continuous variables, such as temperature, wind speed or global radiation,  quantile forecasts  allow one to work with a continuous forecast as the original one by defining a nominal probability level. The choice of a probability level is directly related to the user's loss function: a quantile forecast at a given probability level is the optimal forecast for users with a specific asymmetry in their loss function \citep{koenker99,fh07,gneiting2011b}.

Based on the relationship between user's loss function and quantile forecast level, the quantile score (QS) is the natural scoring rule for assessing the quality of quantile forecasts \citep{koenker99,fh07,gneiting2011b}. More recently, the verification of quantile forecasts has benefited from the tradition and concepts stemming from the probability forecast verification framework. It has been shown that QS is a \textit{proper} scoring rule and a decomposition of the score has been proposed \citep{Bentzien2014}. The QS decomposition provides information about \textit{reliability} and \textit{resolution}, two other fundamental attributes of a probabilistic forecast \citep{toth2003}. 

The aim of the paper at hand is to extend the range of verification methods dedicated to the assessment of quantile forecasts. In particular, the assessment of quantile forecasts from the user's perspective, in a decision-making  framework, is explored here. Based on a simple cost-loss model, the concepts of  forecast discrimination and forecast value are revisited focusing on a specific user rather than on an specific event. First, a new tool is proposed for the analysis of user-based discrimination. The so-called relative user characteristic (RUC) curve and the associated summary measure are shown to be adequate for the assessment of quantile forecast discrimination ability. Secondly, quantile forecast value is discussed as an application of the value score to quantile forecasts. The quantile value plot, showing the economic value of a forecast as a function of a range of events of interest, is proposed as a new tool for the visualization of quantile forecast performance. Finally, the relationship between quantile forecast value and quantile skill score is discussed in the same vein as the relationship between probability forecast value and Brier skill score \citep{murphy69}. The concepts developed are first illustrated with the help of synthetic datasets and in a second step applied to probabilistic forecasts derived from an EPS.

The manuscript is organized as follows: Section \ref{sec:data} describes the datasets that are used to illustrate the discussion.
Section \ref{sec:def}  introduces definitions and notations  and describes the relationship between quantile forecast and forecast user within a cost-loss model framework. Section \ref{sec:discr} discusses the concept of discrimination  and Section \ref{sec:value} the application of the economic value score to quantile forecasts. Section \ref{sec:conc} presents the conclusions.

\section{ Data}
\label{sec:data} 

\subsection{Synthetic datasets}
\label{subsec:toy}
In order to illustrate the concepts discussed hereafter, we make use of synthetic and real datasets. The synthetic data are derived from a toy-model based on normal distributions often used to illustrate verification discussions \citep[e.g.][]{hamill2001,weigel2011}. The toy-model is kept simple  in order to facilitate the interpretation of the results. 

We consider a signal \(s\), normally distributed, written \(s \sim \mathcal{N}(0,1)\).  We assume that the 
observations are randomly drawn from a distribution $\mathcal{N} (s,1)$ and the associated predictive distribution described by $\mathcal{N} (s+\beta,\sigma)$
where \(\beta\) is the unconditional bias parameter and \(\sigma\) the dispersion parameter. We define the following test-cases: 
\begin{enumerate}
    \item[\emph{\(A_0\)}]: \(\beta=0\), \(\sigma=1\) (a perfect probabilistic forecast) ,
    \item[\emph{\(A_1\)}]: \(\beta=-0.75\), \(\sigma=1\) (a biased forecast), 
    \item[\emph{\(A_2\)}]: \(\beta=0\), \(\sigma=1/3\) (an underdispersive forecast),
    \item[\emph{\(B\)}]:  \(\beta=\epsilon_B\), \(\sigma=1\) (a forecast with white noise),
\end{enumerate}
where  \(\epsilon_B\) is derived from a uniform distribution defined on \(]-5,5[\). The  first three datasets \(A_0\), \(A_1\) and \(A_2\)  differ only in terms of biases while the fourth dataset \(B\) corresponds to a forecast with a dynamically disturbed signal.

\subsection{COSMO-DE-EPS}
\label{subsec:cdeps}
Real datasets are provided by COSMO-DE-EPS, a regional ensemble prediction system run operationally at Deutscher Wetterdienst, Offenbach, Germany. The ensemble system is based on a 2.8 km grid resolution version of the COSMO model \citep{stepp03,bald11} 
with  a model domain that covers Germany and parts of the neighbouring countries. The ensemble comprises  20 members including variations in initial conditions, physics parameterisations and boundary conditions \citep{gtpb10,pbtg12}. 

COSMO-DE-EPS has been first developed focusing on high-impact weather events \citep{bbtg2013,zbb2013} and is planned to be used for energy-applications. The focus in this paper is on global radiation which is the main weather variable affecting solar energy forecasts. 
Verification is applied to the 0300UTC run with a forecast horizon ranging between  5 and 15 hours. Two periods of 3 months  are compared: winter (December, January, February) 2012/2013 and summer (June, July, August) 2013. The observation dataset consists of pyranometer measurements from 32 stations distributed over Germany and quality controlled \citep{becker2012}. 

Global radiation forecasts and observations are transformed into clearness index before verification. The clearness index is defined as the ratio between global radiation at ground and global radiation at the top of the atmosphere \citep{badescu}. This pre-processing of the data allows climatological effects and misinterpretation  of the verification results to be avoided \citep{hamilljuras2006}.

\section{Definitions and framework}
\label{sec:def} 

\subsection{Quantile forecast, quantile score, and quantile skill score}
We first consider the quantity to be forecast (or \textit{observation}) \(\Omega \in \Re\)  that we assume to be  a continuous random variable driven by a stochastic process. An observed event \(E\) is defined by a threshold  \(\omega\) as \(E :\Omega \ge \omega\).  The base rate \(\pi\) of an event \(E\) (or climatological frequency) corresponds to:
\begin{equation}
\pi = Pr( \Omega \ge \omega).
\label{equ:pi}
\end{equation}

\begin{figure}
\centering
\includegraphics[width=6cm]{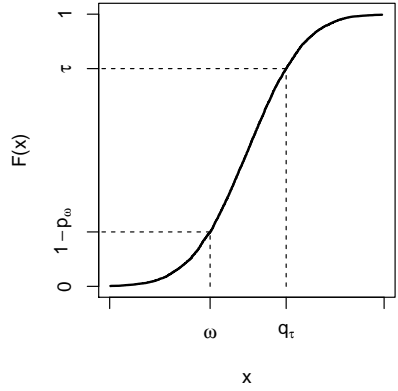}
\caption{
Example of a predictive cumulative distribution function  $F(x)$. Probabilistic products are derived either fixing a threshold \(\omega\) and deriving the associated probability forecast \(p_\omega\), or fixing a probability level \(\tau\) and deriving the associated quantile forecast \(q_\tau\).
}
\label{fig:cdf}
\end{figure}

Consider now a   predictive cumulative distribution  \(F(x)\). 
The  probability  forecast \(p_\omega\) of event \(E\) is defined as:
\begin{equation}
p_\omega= 1- F(\omega). 
\label{equ:prob}
\end{equation}
The quantile forecast \(q_\tau\) at probability level \(\tau\) ($0 \le \tau \le 1$) is defined as: 
\begin{equation}
q_\tau := F^{-1}(\tau)=\text{inf}\{y:F(x)\ge \tau \}
\label{equ:cdf}
\end{equation}
such the relationship between a probability forecast and a quantile forecast is expressed as: 
\begin{equation}
p_{q_\tau}=1-\tau.
\label{equ:pq}
\end{equation}
Figure \ref{fig:cdf} shows an example of a cumulative distribution function \(F(x)\). A threshold  \(\omega\) and  the associated probability forecast  \(1-p_\omega\) as well  as a probability level \(\tau\) and the associated quantile forecast  \(q_\tau\) are shown on the plot.

The quantile score (QS) is the scoring rule applied in order to assess the quality of a quantile forecast. QS is based on an asymmetric piecewise linear function \(\rho_\tau\) called the check function. The check function was first defined in the context of quantile regression \citep{koenker78}: 
\begin{equation}
 \rho_{\tau}(u) = u[\tau - I(u<0)] =
\left\lbrace
\begin{array}{ccc}
\tau u  & \mbox{if} & u \ge 0\\
(\tau-1)u  & \mbox{if} & u < 0\\
\end{array}\right.
\label{equ:check}
\end{equation}
where $I(.)$ is an indicator function having value 1 if the condition in parenthesis is true and zero otherwise.
QS results from the mean of the check function applied to the pairs \(i=1,...,N\) of observation \(\Omega_i\) and quantile forecast \(q_{\tau,i}\) following 
 \begin{equation}
 QS =  \dfrac{1}{N}  \sum_{i=1}^{N} {\rho_\tau(\Omega_i-q_{\tau,i})},
 \label{equ:QS}
\end{equation}
where \(N\) is  the size of the verification sample. Developing Eq. \eqref{equ:QS} we can write
 \begin{equation}
 QS = \dfrac{1-\tau}{N} \sum_{i:\Omega_i < q_{\tau,i}} ( q_{\tau,i} - \Omega_i) 
+    \dfrac{\tau}{N} \sum_{i:\Omega_i \geq q_{\tau,i}} (\Omega_i  - q_{\tau,i} )
\label{equ:QSf}
\end{equation}
The scoring rule consists of penalties per unit \(1-\tau\) and \(\tau\) associated with under-forecasting and over-forecasting, respectively.

Skill scores are computed in order to measure the relative benefit of using a forecast compared to a reference forecast \citep{wilks}. The quantile skill score (QSS) measures the skill of a quantile forecast  compared to a reference quantile forecast. Considering the climatology as reference, QSS corresponds to:
 \begin{equation} 
QSS=  \dfrac{QS_\text{forecast} -QS_\text{climate}}{QS_\text{perfect}-QS_\text{climate}} = 1-\dfrac{QS_\text{forecast}}{QS_\text{climate}}    
\label{equ:QSS}
\end{equation}
where  \( QS_\text{forecast}\), \( QS_\text{perfect}\) and \(QS_\text{climate}\) represent the quantile scores of the forecast under assessment, of a perfect deterministic forecast and of a climatological \(\tau\)-quantile forecast, respectively.  \( QS_\text{perfect}\), by definition, equals 0 and a climatological \(\tau\)-quantile forecast, noted \(\Omega_\tau\), is here defined as the \(\tau\)-quantile of the observation distribution over the verification sample.

\subsection{Cost-loss model and optimal decision-making}
\label{sec:costmodel}
The framework used to discuss the concept of \textit{use}r and \textit{decision-making} is based on a static cost-loss model \citep{thompson62,katz1997}.  The cost-loss  model  describes situations of dichotomous decisions: a user has to decide whether or not to take protective action against potential occurrence of an event \(E\). The decision is made based on a decision variable (or forecast)  \(\Lambda\). A decision criterion \(\lambda\) applied to the decision variable defines an action \(A: \Lambda \ge \lambda\). Taking action implies a cost \(C\). In the case of occurrence of the event \(E\) without preventive action, a loss  \(L\)  is encountered. The cost-loss ratio is denoted \(\alpha\): 
\begin{equation} 
  \alpha=\dfrac{C}{L}.
  \label{equ:alpha}
\end{equation}
A user with cost-loss ratio \(\alpha\) is called hereafter an \(\alpha\)-user. Based on this simple model the optimal decision strategy  of an \(\alpha\)-user can be discussed \citep[e.g.][]{Richardson2011}. The problem consists of finding, for a decision variable \(\Lambda\), the \textit{critical} decision criterion \(\lambda_\alpha\) that minimizes the \(\alpha\)-user mean expense if actions are taken when \(\Lambda \ge \lambda_\alpha\).

Consider first the case of a probability forecast \(p_\omega\) as a decision variable. Based on \(p_\omega\), does the user have to take action or not? In order to answer this question, the average expenses in the cases of positive and negative answers are compared. If the answer is yes, the user encounters a cost  \(C\) on every occasion, so the average expense \(\bar{E}_\text{yes}\) is simply
\begin{equation} 
  \bar{E}_\text{yes}=C.
  \label{equ:yes}
\end{equation}
If the answer is no, the user has no cost but a loss \(L\) on each occasion where the event occurs, so on average
the user's expense \(\bar{E}_\text{no}\) is 
\begin{equation} 
   \bar{E}_\text{no}=L   Pr(\Omega \geq \omega \mid p_\omega),
\label{equ:no}
\end{equation}
where \(Pr(\Omega \geq \omega \mid p_\omega)\) is the probability that the event occurs when the probability forecast \(p_\omega\) is issued. 
So, users with a cost-loss ratio \(\alpha < Pr(\Omega \geq \omega \mid p_\omega)\) should take preventive action, while users with a greater cost-loss ratio should not.
The critical decision criterion  \(p_\omega^\star\) associated with the decision variable \(p_\omega\) is thus defined as 
\begin{equation} 
  p_\omega^\star = \{p_\omega \ | \Pr(\Omega \geq \omega \mid p_\omega) = \alpha \}.
\label{equ:optdec}
\end{equation}
Thus, the action based on the probability forecast  \(A: p_\omega \geq p_\omega^\star\) optimizes the user's mean expense in the long term.  

If the forecast is reliable, we have by definition \( \Pr(\Omega \geq \omega \mid p_\omega) =  p_\omega\): the event actually happens with an observed relative frequency consistent with the forecast probability \citep{brocker2009}. The optimal decision is then to take action if 
\begin{equation}
p_\omega \geq \alpha.
\label{equ:optp}
\end{equation}
When the probability forecast is compared to the cost-loss ratio in order to decide whether or not to take action  (without additional information about forecast reliability), we say that the probability forecast is taken at face value. For example, consider users who have to decide whether or not to take preventive action against precipitation occurrence. If the forecast probability of precipitation is 10\%,  users with cost-loss ratio lower than 10\% take action. If the forecast is not reliable, the critical decision criterion is no longer \(\alpha\) but has to be adjusted following Eq. \eqref{equ:optdec}. Statistical adjustments of the forecast based on past data is usually referred as \textit{forecast calibration} \citep[e.g.][]{gneit2007}.

Consider now  a quantile forecast \(q_{\tau}\) as a decision variable. We apply the same reasoning as for a probability forecast. The critical decision criterion \(q_\tau^\star\) associated with \(q_{\tau}\) is defined as
\begin{equation}
q_\tau^\star = \{ q_\tau \ | \ Pr( \Omega \geq \omega  \mid  q_\tau ) =  \alpha  \}
\label{equ:cdcq}
\end{equation}
such that taking action when \(q_\tau \geq q_\tau^\star \) minimizes the user mean expense.  By definition,  a quantile forecast is reliable if it satisfies 
\begin{equation}
Pr( \Omega \geq \omega \mid   q_\tau = \omega ) =  1-\tau ,
\label{equ:qrelia}
\end{equation}
i.e.\ the observed relative frequency of the event defined by the quantile forecast  is consistent with the quantile forecast probability level. Eq. \eqref{equ:cdcq} has a straightforward solution
\begin{equation} 
q^\star_\tau = \omega
\label{equ:optq}
\end{equation}
 when the decision variable is the quantile forecast at probability level \(\tau\) defined as 
\begin{equation}
\tau = 1-\alpha.
\label{equ:taualpha}
\end{equation}
Taking action when \(q_\tau \geq \omega\) with \(\tau= 1-\alpha\) is equivalent to taking action when \(p_\omega \geq \alpha\) since the cumulative probability distribution function \(F(x)\) is by definition monotonically increasing (see e.g. Figure \ref{fig:cdf}). Hence, a quantile forecast is taken at face value when the user's decision is made based on the comparison of the forecast with the event threshold \(\omega\). In our example, if the  90\%-quantile forecast of precipitation  is greater than zero, a user with cost-loss ratio \(\alpha=1-0.9=0.1\) takes preventive action.

In a general form, the critical decision criterion \(\lambda_\alpha\) for an \(\alpha\)-user is defined by 
\begin{equation}
\lambda_\alpha = \{ \lambda \ | \ Pr( \Omega \geq \omega  \mid  \lambda ) =  \alpha  \}
\label{equ:cdc}
\end{equation}
where  the decision variable could  equally be the probability forecast \(p_\omega\)  or the quantile forecast \(q_\tau\) with \(\tau=1-\alpha\). Provided that the forecasts are reliable, the critical decision criteria are known and have a simple expression (Eqs (\ref{equ:optp},\ref{equ:optq})). In the following, we say that the decision variable is taken at \textit{face value} when the user applies the decision criterion valid for a reliable forecast, irrespective of whether the forecast is actually reliable or not.

\subsection{Quantile forecast user}
\label{subsec:user}

\begin{figure}
\centering
\includegraphics[width=9cm]{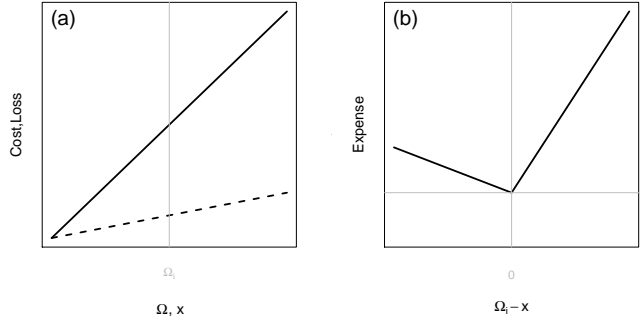}
\caption{ 
    (a) Cost (dashed line) as a function of the level of protection  \(x\) and loss (full line) as a function of the observation \(\Omega\). An observation \(\Omega_i\) is represented by a vertical line.
    (b) Expense as a function of the difference between the observation \(\Omega_i\) and the level of protection \(x\). The horizontal line indicates the expense for a perfect level of protection.
}
\label{fig:CLEx}
\end{figure}

The dichotomous decision problem is extended to a continuous decision problem considering the cost \(C\)  and  the loss \(L\) as unitary cost and unitary loss, respectively \citep{epstein69,roulston2003}. The cost of taking protection is a linear function of the level of protection \(x\) and  the loss without protection is a linear function of the observation \(\Omega\), as illustrated in Figure \ref{fig:CLEx}. The optimization problem consists of finding the level of protection that minimizes the expected user expense. 
 
Considering a variable defined on \(\Re+\)  (the generalization to variables defined on \(\Re\) is straightforward), the expense associated with a level of protection \(x\) corresponds to \(C  x\). If the observation is \(\Omega\), then protection is perfect if $x = \Omega$. But if $x>\Omega$, then there is an unnecessary expense due to a larger  level of protection than is actually needed. If the observation \(\Omega\) is greater than the level of protection, then additionally a loss \(L  (\Omega- x)\) is encountered. Formally, we can write the expense function $E$ as 
\begin{equation}
E =
\ \begin{cases}
C   (x -\Omega)   & \text{if } \Omega < x \\
(L -C)  (\Omega- x) & \text{if } \Omega \geq x.
  \end{cases}
\label{equ:expi}
\end{equation}
The expense function is represented in Figure  \ref{fig:CLEx}.
If divided by $L$, the expense function is an asymmetric loss function equivalent to the check function defined in Eq. \eqref{equ:check}, where the asymmetry is given by \(\tau = \frac{L-C}{L}\). Thus the optimal level of protection \(x^\star\) which minimizes the user's mean expense corresponds to the $1-\alpha$ quantile of the true predictive distribution of $\Omega$.

This result is not new: quantile forecasts arise as an optimal solution for users with an asymmetric linear loss function \citep{koenker78,chris97}. More recently, it has been shown that quantile forecasts are optimal forecasts in a stochastic optimization framework for a more general class of loss functions \citep{,gneiting2011}. 

Asymmetric loss functions find a number of applications, in particular for operational decision-making problems related to the integration of renewable energies into the electricity grid. For example, asymmetric loss functions can be associated with market participants who want to optimize their bids or system operators who have to optimize their reserves. The user's optimal forecast corresponds then to a specific quantile of the predictive distribution where the probability is defined by the user's cost-loss ratio \citep{pinson2007,pinson2013}.

\section{Discrimination}
\label{sec:discr} 

Based on the discussion developed in the previous Section, continuous decision making is seen in the following as a \textit{continuum} of dichotomous decisions. For each threshold \(\omega\) of the event spectrum, the question is whether to take action for the next unit of the variable. The adequate decision for a user in order to minimize the expected expense is a function of his (her) cost-loss ratio as defined in Eq. \eqref{equ:cdc}. Moreover, the relationship between cost-loss ratio and quantile probability level, \(\tau=1-\alpha\), makes implicit the cost-loss ratio  \(\alpha\) of a user as soon as the level \(\tau\) of the quantile forecast used as decision variable is selected. 

\subsection{General verification framework}
A general framework for forecast verification is based on the joint distribution of forecasts and observations \citep{murphy87}. The overall agreement between forecasts and observations is called quality and is measured by scoring rules, like QS for quantile forecasts. In order to access more information about the forecast performance, two factorizations of the joint distribution, into conditional and marginal distributions, can be applied: the \textit{calibration-refinement} (CR) factorization when conditioning on the forecasts  and  the \textit{likelihood-base rate} (LBR) factorization when conditioning on the observations. Summary measures based on these two factorizations are associated with attributes, fundamental characteristics of the forecast. Reliability and resolution are derived from the CR  factorization while discrimination is derived from the LBR factorization \citep{murphy92}.

Here the focus is on discrimination, the key forecast attribute for decision-making processes. A general definition of discrimination is "the ability of a forecasting system to produce different forecasts for those occasions having different realized outcomes" \citep{wilks}. Discrimination assessment is discussed in terms of \textit{event} and \textit{action} within the dichotomous decision framework. Regarding the LBR factorization, it is common practice to analyse discrimination in terms of hit rate \(H\) and false alarm rate \(F\) defined as  
\begin{equation}
    H  =  Pr( \Lambda \ge \lambda \mid \Omega \ge \omega )
    \label{equ:H}
\end{equation}
and
\begin{equation}
    F  =  Pr( \Lambda \ge \lambda \mid \Omega < \omega),
    \label{equ:F}
\end{equation}
respectively. Actions \(A: \Lambda \ge \lambda \) and events \(E:\Omega \ge \omega\) are dichotomous, each presenting  two alternatives, so H and F can be easily derived from the construction of a \(2\times2\) contingency table. No discrimination corresponds to the case where: 
\begin{equation}
    H =  F 
    \label{equ:HFlambda}
\end{equation}
for all \(\lambda \in \Lambda\) and \(\omega \in \Omega\), meaning that actions and event occurrence are independent \citep{brocker2014}. 

\subsection{Event-based discrimination}
\begin{figure}[htb]
\centering
\includegraphics[width=5cm]{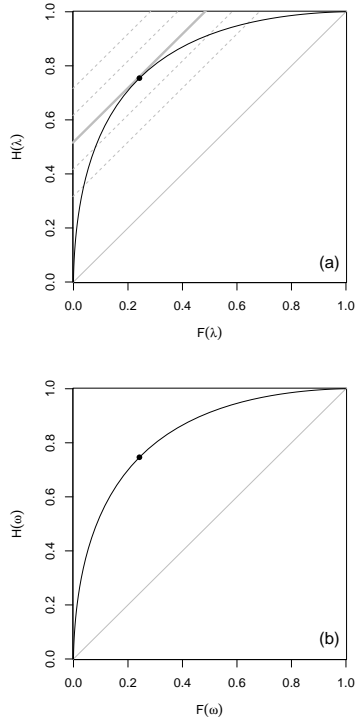}
\caption{
    Discrimination curves for decision variables from the synthetic dataset \(A_0\). The diagonal lines are the no discrimination lines. The points correspond to the (\(F\),\(H\)) pair for the event \(\Omega \geq 0\) and the action associated with the  $50\%$-users. 
    (a) ROC curve of the probability forecast \(p_{0}\) for the event  \(E: \Omega \geq 0\), with base rate \(\pi=0.5\), and equi-cost lines (in grey) of slope \(\gamma =1\). 
    (b) RUC curve of the quantile forecast \(q_{0.5}\) for the user with cost-loss ratio \(\alpha=0.5\).
}
\label{fig:roccas}
\end{figure}

We first focus on one particular event defined by a threshold \(\omega\), with event-specific  hit rate \(H_\lambda\) and false alarm rate \(F_\lambda\). A popular way to assess discrimination (Eq. \eqref{equ:HFlambda}) is to plot the set of points (\(F_\lambda,H_\lambda\)) for a range of actions with \(\lambda \in \Lambda\). The resulting curve is known as the relative operating characteristic (ROC) curve. When action and event occurrence are independent, the ROC curve is a diagonal line. Concavity of the curve indicates a discrimination ability in the forecast and the area under the curve (AUC) becomes a quantitative measure of forecast discrimination \citep{mason82}. Figure  \ref{fig:roccas} (a) shows an example of a ROC curve for the synthetic dataset \(A_0\). The event of interest is \(E: \Omega \geq 0\) with a base rate $\pi = Pr( \Omega \geq 0)$ of  0.5.  The respective forecast probability \(p_0 =1-F(0)\) is used as decision variable.

The interpretation of the ROC curve can be related to the dichotomous decision model described in Section \ref{sec:costmodel} as discussed for example in \cite{Richardson2011}. In order to describe this relationship, we consider the slope of the ROC curve, defining first the gradient of a line joining two successive ROC points
 (\(F_{\lambda},H_{\lambda}\)) and (\(F_{\lambda+\Delta \lambda},H_{\lambda+\Delta \lambda}\)):
\begin{equation}
\dfrac{H_{\lambda}-H_{\lambda+\Delta \lambda}}{F_{\lambda}-F_{\lambda+\Delta \lambda}} =
\dfrac
{ Pr( \Lambda \ge \lambda \mid \Omega \ge \omega ) -  Pr( \Lambda \ge \lambda+\Delta \lambda \mid \Omega \ge \omega )}
{ Pr( \Lambda \ge \lambda \mid \Omega < \omega ) -  Pr( \Lambda \ge \lambda+\Delta \lambda \mid \Omega < \omega )}.
\end{equation}
The slope of the curve \(\gamma\)  is obtained when  \(\Delta \lambda\) tends to \(0\):
\begin{equation}
\gamma(\lambda,\omega)  = \dfrac{Pr( \Lambda = \lambda \mid \Omega \ge \omega) } {Pr( \Lambda = \lambda \mid \Omega < \omega) } 
\label{equ:LR1}
\end{equation}
where the ratio is also know as the \textit{likelihood ratio} \citep{brock2011}. Using the Bayes rule and the definition of the critical decision criterion of an \(\alpha\)-user in Eq. \eqref{equ:cdc}, we can write
\begin{equation}
\gamma(\lambda_\alpha,\omega) = \dfrac{1-\pi}{\pi} \dfrac{\alpha}{1-\alpha} 
\label{equ:LR2}
\end{equation}
where \(\pi = Pr(\Omega \ge \omega)\) is the base rate of an event \(E:\Omega \ge \omega\) and \(\lambda_\alpha\) the corresponding critical decision criterion of an \(\alpha\)-user. 

The range of decision criterion \(\lambda\) used to derive the ROC curve (\(F_{\lambda},H_{\lambda}\)) corresponds  to a range of critical decision criteria associated with users with different cost-loss ratios. Each point of the ROC curve is associated with a specific \(\alpha\)-user that is identified by the slope of the curve at that point. The slope possibly  ranges between \(0\) and \(+\infty\) at the right-top and the bottom-left corners of the ROC plot respectively. Moving along the curve from the top to the bottom consists in varying the cost-loss ratio  \(\alpha\) between \(0\) and \(1\).

For example, consider a user with a cost-loss ratio \(\alpha= 50\%\). In Figure \ref{fig:roccas}, the point of the ROC curve  with slope \(\gamma= 1 \) is highlighted (\(\alpha =0.5,\pi=0.5\) in Eq. \eqref{equ:LR2}). This point indicates the performance of the forecast in terms of \(H\) and \(F\) for this particular user.  Conversely, the decision criterion applied  to obtain this point corresponds to the critical decision criterion for the 50\%-user.

The ROC curve applied to a  decision variable, then,  corresponds to testing whether actions and event occurrence are independent for one event and  a range of users with different cost-loss ratios. The ROC curve is an \textit{event specific} but \textit{user unspecific} discrimination tool and is therefore well-adapted to probability forecast discrimination assessment. 

\subsection{User-based discrimination}
We focus now on a user with cost-loss ratio \(\alpha\). The critical decision criterion \(\lambda_\alpha\) defines the action of this specific user with respect to an event. We define then the user-specific hit rate \(H_\omega\)  and false alarm rate \(F_\omega\)  as in Eqs \eqref{equ:H} and \eqref{equ:F} for a fixed \(\alpha\). In order to test Eq. \eqref{equ:HFlambda}, the set of points (\(F_\omega,H_\omega\)) are plotted for a range of events. We call the resulting curve a \textit{relative user characteristic} (RUC) curve because it is a comparison of two user characteristics (\(F_\omega\) and \(H_\omega\)) as the event definition varies. As for the ROC curve, the no discrimination line corresponds to the  diagonal line and concavity of the curve indicates forecast discrimination ability.

Figure \ref{fig:roccas} (b) shows an example of a RUC curve valid for a user with cost-loss ratio \(\alpha=50\%\). In this example, the decision variable is the 50\%-quantile forecast from the synthetic dataset \(A_0\). Moving along the RUC curve from the bottom left corner to the top right corner involves varying the event under focus, the event's base rate varying from \(0\) to \(1\), respectively. The point with slope \(\gamma=1\) corresponds to the event \(E:\Omega \geq 0\) with base rate \(\pi=0.5\). This point is obviously the same as in Figure \ref{fig:roccas} (a). 

In order to produce a RUC curve, critical decision criteria have to be known for a range of events. They can be estimated resolving Eq. \eqref{equ:cdcq} numerically.  In practice,  critical decision criteria can also be estimated by means of a reliability diagram. For example, a reliability diagram for quantile forecasts plots the conditional observed quantile as a function of quantile forecast categories \citep{Bentzien2014}. With regard to Eq. \eqref{equ:qrelia}, we can deduce that the mean forecast in each forecast category  (horizontal axis of the diagram) is an estimation of the critical decision criteria associated with the events defined by the corresponding conditional observed quantile (vertical axis of the diagram).

The RUC curve is user specific (and event unspecific) and therefore well-adapted to quantile forecast discrimination. A summary measure of quantile discrimination ability is obtained mimicking the ROC framework: the area under the RUC curve, noted here \(AUC^\prime\), is proposed as a quantitative measure of discrimination for quantile forecasts. Considering \(n_E\) events \(E_i: \Omega \geq \omega_i\), \(i=1,...,n_E\) with increasing base rate,  \(AUC^\prime\) is estimated by a trapezoidal approximation as
\begin{equation}
    AUC^\prime = \sum_{i=0}^{n_E}  0.5  ( H_{\omega_{i+1}} + H_{\omega_{i}}) ( F_{\omega_{i+1}} - F_{\omega_{i}}) 
    \label{equ:aucprime}
\end{equation}
with the trivial points \(H_{\omega_{0}}=F_{\omega_{0}}=0\) (for an event of base rate 0) and \(H_{\omega_{n_E+1}}=F_{\omega_{n_E+1}}=1\) (for an event of base rate 1). In order to reduce the biases introduced by the limited number of RUC points, the RUC curve can be fitted  under a bi-normal assumption. The procedure involves considering \(F_\omega\) and \(H_\omega\) as both expressed as integrations of the standard normal distribution \citep{mason82}. The bi-normal model has been shown to be valid in most cases when applied in the ROC framework \citep{mason2002,atger2004}.

\begin{figure*}[thb]
\centering
\includegraphics[width=17cm]{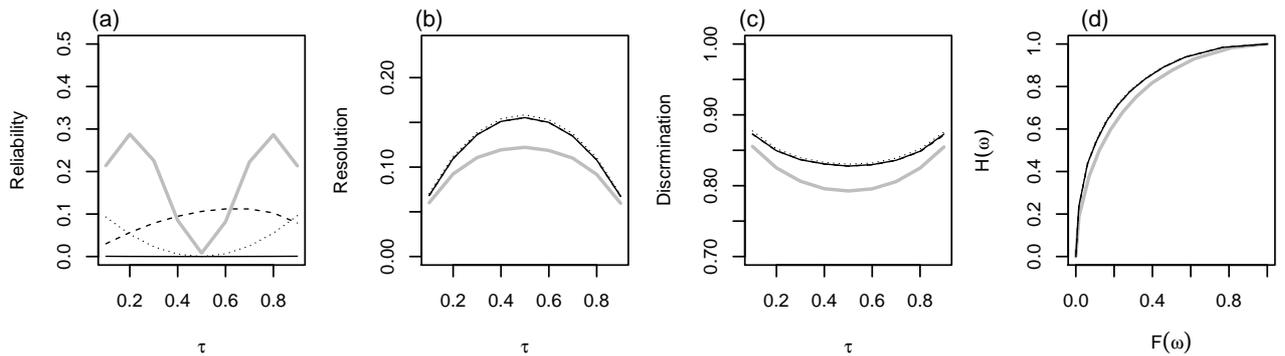}
\caption{ 
    (a) Reliability, (b) resolution and (c) discrimination as a function of the probability level \(\tau\)  of the \(\tau\)-quantile forecasts and (d) RUC curves for the 50\%-quantile forecasts ($\tau=0.5$). The results are shown for the simulation test cases \emph{\(A_0\)} (full lines),  \emph{\(A_1\)} (dashed lines), \emph{\(A_2\)} (dotted lines) and \emph{\(B\)} (full grey line).
}
\label{fig:ABCDE}
\end{figure*}

The properties of the RUC curve and  \(AUC^\prime\) are discussed with the help of illustrative examples based on 4 simple simulation test cases (see Section \ref{subsec:toy}). In Figure \ref{fig:ABCDE}, the forecast attributes reliability, resolution and discrimination  are shown as a function of the probability level  \(\tau\) of the  \(\tau\)-quantile forecast under assessment.  RUC curves for the 50\%-quantile forecasts are also shown. Quantile forecast reliability and resolution are estimated using the decomposition of the quantile score \citep{Bentzien2014}  while discrimination curves and summary measures are estimated based on the bi-normal assumption.

Figures \ref{fig:ABCDE} (a) shows the lack of reliability, which occurs by construction in the simulations \(A_1\), \(A_2\) and \(B\). In Figures \ref{fig:ABCDE} (b) and \ref{fig:ABCDE} (c), resolution and discrimination measures deliver a similar message comparing the different simulations which illustrates the idea that "resolution and discrimination are the two faces of the same coin" \citep{brocker2014}. Resolution and discrimination exhibit however different behaviours as a function of the probability level reflecting  the fact that the first takes the forecaster's perspective and the second the user's perspective. Moreover, discrimination ability is  identical for the simulations \(A_0\),  \(A_1\) and \(A_2\): they are unaffected by biases and dispersion errors. Indeed, \(AUC^\prime\) is by construction insensitive to conditional and unconditional biases. In contrast, the forecast derived from simulation \(B\) with a perturbed signal presents less discrimination ability than forecasts from the other simulations, in particular for the 50\%-quantile forecast. Focusing on users with cost-loss ratio \(\alpha=0.5\) ($\tau=0.5$) , RUC curves for the 50\%-quantile forecasts of simulations  \(A_0\), \(A_1\), \(A_2\), and \(B\) are shown in Figure \ref{fig:ABCDE} (d). The largest discrepancies between simulations $A$ and $B$ are visible at the centre of the RUC curves, so for events with intermediate base rates, while for events with small or large base rates the RUC curves tend to overlap. 

\section{ Value of quantile forecasts}
\label{sec:value} 

\subsection{Economic value}
The cost-loss model described in Section \ref{sec:costmodel} has been used to develop the concept of economic value of a probabilistic forecast. The forecast value is assessed considering decision-making made by an \(\alpha\)-user about the occurrence of an event. The value of a forecast (also called value score or relative value) is defined as
\begin{equation} 
V = \dfrac{\bar{E}_\text{climate}- \bar{E}_\text{forecast}} {\bar{E}_\text{climate}- \bar{E}_\text{perfect}},
\label{equ:Vdef}
\end{equation}
where the mean expense \(\bar{E}\) of an \(\alpha\)-user is estimated when decisions are based on a forecast (\(\bar{E}_\text{forecast}\)), on a perfect deterministic forecast (\(\bar{E}_\text{perfect}\)), or on climatological information (\(\bar{E}_\text{climate}\)) \citep{richardson2000,wilks2001,zhu02}.  \(V\) is a measure of the  economic gain (or reduction of mean expense) when using a forecast relative to the gain when using a perfect deterministic forecast.
 
Following \textit{e.g.} \cite{Richardson2011}, the mean expense of a forecast user can be written as
\begin{equation} 
\bar{E}_\text{forecast} = F(1-\pi)C -H\pi(L-C)+ \pi L,
\label{equ:Efct}
\end{equation}
where \(H\) and \(F\) are the hit rate and false alarm rate as defined in Eqs \eqref{equ:H} and \eqref{equ:F}, respectively, and \(\pi\) the base rate of the event of interest. A user with a perfect deterministic forecast at hand has to face costs only. The user mean expense corresponds in this case to:
\begin{equation} 
\bar{E}_\text{perfect} = \pi C.
\label{equ:Eperf}
\end{equation}
For a user who bases his (her) decision on climatological information, the optimal mean expense is expressed as 
\begin{equation} 
\bar{E}_\text{climate} = \left\lbrace
 \begin{array}{ccc}
C  & \mbox{if} & \alpha < \pi\\
\pi L  & \mbox{if} & \alpha \ge \pi,\\
\end{array}\right.
\label{equ:Eclim}
\end{equation}
depending on the relationship between cost-loss ratio and base rate.
Combining Eqs \eqref{equ:Efct}-\eqref{equ:Eclim}, 
the value of a forecast can finally be written as:
\begin{equation}
V=\left\lbrace
 \begin{array}{ccc}
(1-F) -\left(\dfrac{\pi}{1-\pi} \right)\left(\dfrac{1-\alpha}{\alpha} \right) (1-H)& \mbox{if} & \alpha < \pi \\
H -\left(\dfrac{1-\pi}{\pi} \right)\left(\dfrac{\alpha}{1-\alpha} \right) F & \mbox{if} & \alpha \geq \pi.  \\
 \end{array}\right.
\label{equ:Vdef2}
\end{equation}
So, the economic value \(V\) is defined for an event with base rate \(\pi\) and a user with cost-loss ratio \(\alpha\).  \(V\) depends on the forecast performance in terms of H and F.

\begin{figure}[htb]
\centering
\includegraphics[width=6cm]{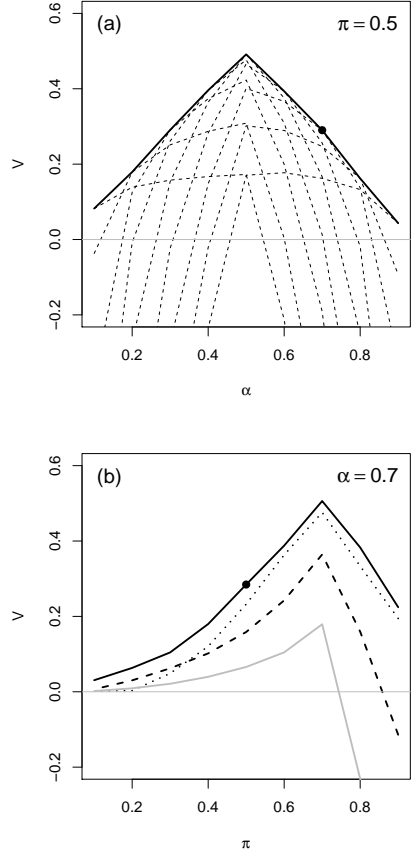}
\caption{
    (a) Value \(V\) of the probability forecast from simulation $A_0$ for the event defined as \(E:\omega \geq 0\) with base rate \(\pi=0.5\). The dashed lines represent the forecast value when the probability levels $0.1,0.2,...,0.9$ are chosen as decision criterion. The full line represents the envelope of the dashed lines.  
    (b) Value \(V\) for users with cost-loss ratio \(\alpha=0.7\) of the \(30\%\)-quantile forecasts taken at face value from the 4 synthetic datasets: \(A_0\) (full black line, square), \(A_1\) (dashed line, triangle), \(A_2\) (dotted line, circle) and  \(B\) (full grey line, cross). The black point is the common point of the two plots: value of the simulation \(A_0\) for the event with base rate $\pi=0.5$ and a user with cost loss ratio $\alpha=0.7$. 
 }
\label{fig:maxval}
\end{figure}

Applied to a probability forecast, the event's base rate is fixed and the value of a probability forecast is generally represented in the form of a probability value plot showing \(V\) as a function of \(\alpha\). An example is provided in Fig. \ref{fig:maxval} (a), applied to simulation $A_0$ considering the event \(E:\omega \geq 0\). The forecast value curves are plotted for a range of probabilities as decision criterion, then the optimal values for each \(\alpha\)-user (the upper envelope of the relative value curves) is selected to represent the value of the probabilistic forecast system \citep[e.g.][]{richardson2000,wilks2001}. The probability value plot is related to the ROC framework since the pairs $(F,H)$ of Eq. \eqref{equ:Vdef2} are the ones used to draw the ROC curve. It has also been shown that the overall value of a probability forecast, considering all potential users, corresponds to the Brier skill score of the forecast if the distribution of cost-loss ratio is uniform over all users \citep{murphy69,Richardson2011}.

\subsection{Quantile value plot }

Applied to a quantile forecast, so focusing on a  \(\alpha\)-user, the value score is evaluated  for a range of events of interest  defined for example by their base rate \(\pi\). A new tool is therefore proposed for the assessment of quantile forecast performance:
the quantile value plot which represents how \(V\) varies as a function of \(\pi\).  This is illustrated in Figure \ref{fig:maxval} (b). The value of the \(30\%\)-quantile forecasts is plotted when the quantile forecasts derived from simulations  \emph{\(A_0\)}, \emph{\(A_1\)}, \emph{\(A_2\)}, and \emph{\(B\)} are taken at face value.
Taking a quantile at face value means using it as it is, so for each event it implies considering the event threshold as decision criterion (see Section \ref{sec:costmodel}). An alternative is to apply the critical decision criteria, i.e. to use the $(F,H)$ pairs from the RUC curve to estimate the value in Eq. \eqref{equ:Vdef2}. We talk then about \textit{potential} value since it corresponds to the maximum value of the forecast, i.e. the maximum that could be potentially reached if an adequate calibration is applied to the forecast. Indeed, value and \textit{potential} value are by definition  identical if the forecast is reliable.

A parallel between probability value plot and quantile value plot can be draw. In a probability value plot, the decision variable is a probability forecast, the base rate  \(\pi\) of the event under focus is fixed and the forecast value \(V\) is then plotted for a range of cost-loss ratios. The role of \(\alpha \) and \(\pi\) are inverted in order to produce a quantile value plot rather than a probability value plot.  The cost-loss ratio is defined by the quantile probability level  and a range of events of interest are scanned. It results that 
the cost-loss ratio of the end-user does not appear explicitly in a quantile value plot as is the case for the value plot for probability forecasts. 

The fundamental properties  of \(V\) are however the same when focusing on one event or on one user. These properties \cite[demonstrations  can be found e.g. in][]{Richardson2011} are recalled here. First, the forecast value reaches its maximum when \(\pi=\alpha\) (or noted differently when  \(\pi=1-\tau\)). For instance, a forecast user with a  cost-loss ratio of \(\alpha = 0.1\) draws a maximum benefit from a forecast if his (her) event of interest has a climatological probability of occurrence of \(10\%\). Secondly, the value of a reliable forecasts (full line in Figure \ref{fig:maxval} (b)) is always greater than the value of the same forecast with biases (dashed and dotted lines in Figure \ref{fig:maxval} (b)). The value of the reliable forecast corresponds to the potential value of the two other datasets. Finally, the potential value is by definition always positive. 

\subsection{A real example}
\label{subsec:cdeps}
 
\begin{figure*}[htb]
\centering
\includegraphics[width=17cm]{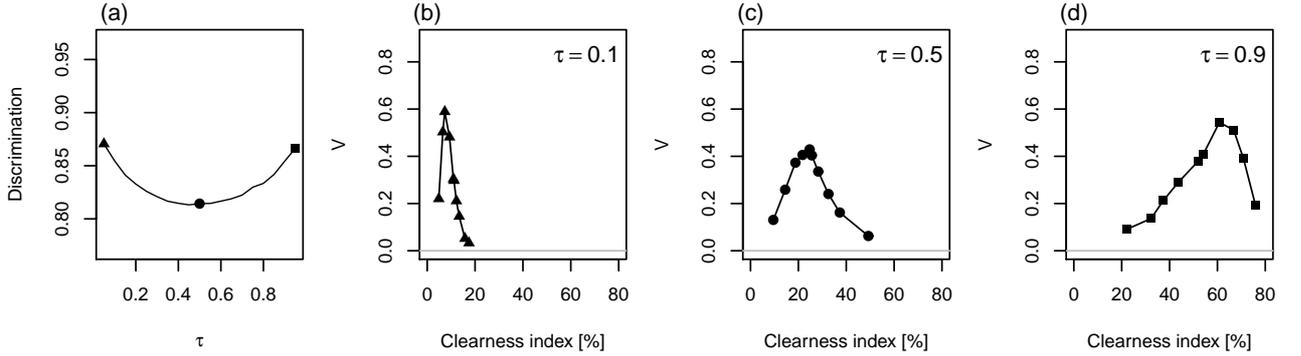}
\caption{ Verification results for COSMO-DE-EPS global radiation forecasts during winter 2012/2013:  
     quantile discrimination ability (\(AUC^\prime\)) as a function of the probability level (a),  
    potential value of the 10\%-quantile forecast (b), 50\%-quantile forecast (c) and 50\%-quantile forecast (d)  as a function of the event of interest defined by thresholds of the clearness index in \%.
}
\label{fig:cdeps1}
\end{figure*}

\begin{figure*}[htb]
\centering
\includegraphics[width=17cm]{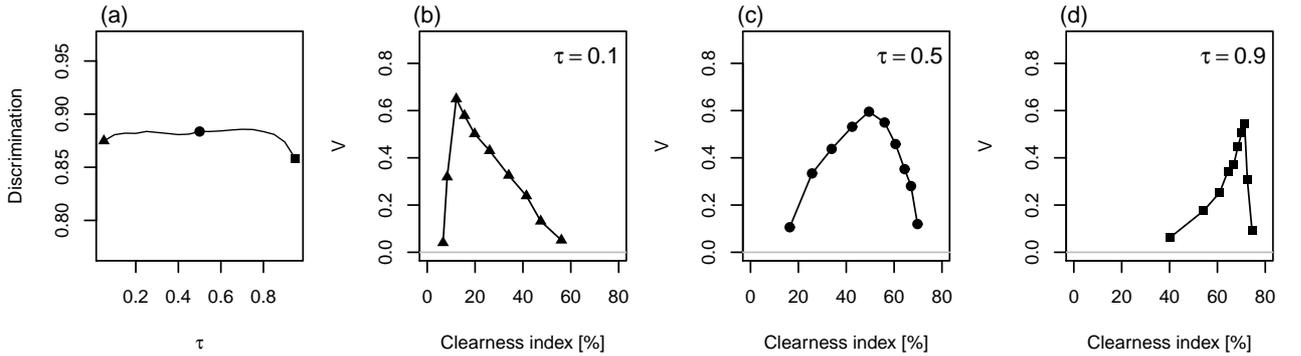}
\caption{ 
    Same as Figure \ref{fig:cdeps1} but for summer 2013.
}
\label{fig:cdeps2}
\end{figure*}

The tools introduced for the assessment of quantile forecast discrimination and value are here applied to a real dataset. Quantile forecasts of global radiation are derived from COSMO-DE-EPS  and assessed for two periods of the year 2013. Results for the winter period are shown in Figure \ref{fig:cdeps1} and results for the summer period in Figure  \ref{fig:cdeps2}. Quantile discrimination is estimated with the area under the RUC curve (\(AUC^\prime\)) for probability levels \(\tau=0.1,0.2,...,0.9\). A deeper analysis is performed for the 10\%-, 50\%- and 90\%-quantile forecasts with the help of quantile value plots.   

The discrimination ability of the EPS quantile forecasts varies as a function of the probability level but is greater than 0.80 which can be interpreted as good performance. For the winter season, discrimination is higher for high and low probability levels than intermediate ones   whereas for the summer season, discrimination is approximately constant over the probability levels with a tendency to decrease for high levels.
Inspection of the quantile value plot allows a deeper insight into the forecast potential performance.  This could be relevant for quantile users with a specific interest in only one part of the event spectrum. The potential value of the quantile forecasts is plotted as a function of event in terms of the clearness index in \% to simplify the reading of the plots. However, an event has a different base rate for each season which complicates a direct comparison of the quantile value plot in Figures \ref{fig:cdeps1} and \ref{fig:cdeps2}.

\subsection{Overall value and Quantile Skill Score}

As a final step in drawing a parallel between probability forecast verification and quantile forecast verification, the relationship between value and skill score with climatology as a reference is explored. It has been shown that the overall value of a probability forecast is equivalent to its Brier Skill Score (BSS) when the users have a uniform distribution of cost-loss ratio \citep{murphy69,Richardson2011}. Similarly, we now investigate  the relationship between the overall value of a quantile forecast  
and its QSS. 

For this purpose, we extend the cost-loss model to more than two observation categories assuming that the cost \(C\) and the loss \(L\)  of the cost-loss model are the unitary increment of cost and loss per unit of variable, respectively, as discussed in Section \ref{subsec:user}. 
Following \cite{Richardson2011}, the overall value is defined as the ratio
\begin{equation} 
V_{all}=  \dfrac{T_{C}- T_{F}} {T_{C}- T_{P}}
\label{equ:Vall}
\end{equation} 
where the total mean expense $T$ of a user is estimated when decisions are based on a climatological forecast ($T_C$), on a perfect deterministic forecast ($T_P$) or on a given forecast ($T_F$) so that Eq. \eqref{equ:Vall} is the extension of Eq. \eqref{equ:Vdef} to all possible events. 

The total expense for a perfect deterministic forecast corresponds to the sum of the costs \(C\) associated with each observation. The total mean expense \(T_P\) can then be expressed as 
\begin{equation} 
T_P= \dfrac{1}{N} \sum_{i=1}^{N} C  \Omega_i .  
\label{equ:TP}
\end{equation}
For a climatological quantile forecast \(\Omega_\tau\), the total expense corresponds to the sum of the costs associated with  \(\Omega_\tau\) and the losses encountered when the observations are greater  than the climatological forecast (\(\Omega_i\geq \Omega_\tau\)). The total mean expense for a climatological forecast \(T_C\) is written as
\begin{equation} 
T_C= \dfrac{1}{N} \sum_{i=1}^{N} C  \Omega_\tau + 
\dfrac{1}{N}  \sum_{i:\Omega_i\geq \Omega_\tau}  L   (\Omega_i- \Omega_\tau ) .
\label{equ:TC} 
\end{equation}
Considering now a sample of quantile forecasts \(q_{\tau,i}\) and the corresponding observations \(\Omega_i\), the total expense of a forecasts user corresponds in that case to the sum of the costs associated with each forecast  \(q_{\tau,i}\) and the losses encountered when  \(\Omega_i \geq q_{\tau,i}\), given by
\begin{equation} 
T_F = \dfrac{1}{N} \sum_{i=1}^N C   q_{\tau,i} + 
 \dfrac{1}{N} \sum_{i:\Omega_i \geq q_{\tau,i}} L  (\Omega_i - q_{\tau,i}) .
\label{equ:TF}
\end{equation} 

Combining Eqs \eqref{equ:TP}-\eqref{equ:TF}, it is shown in the Appendix that the overall value \(V_{all}\) corresponds to QSS (Eq. \eqref{equ:QSS}) with the climatology as a reference based on the assumption of constant cost-loss ratio for all outcomes. 
In other words, extending the dichotomous event-action framework to a continuous framework allows one to turn back to  the `classical` or `natural` measure of performance for quantile forecast. Conversely, using the dichotomous framework provides the keys to making a deeper analysis of the quantile performance at the event level.

\section{Conclusion}
\label{sec:conc} 
Verification measures and tools related to users' decision-making are provided here for quantile forecasts as decision variables. Drawing a parallel with the verification of probability forecasts, the new verification tools allow the scuite of verification methods for quantile forecasts to be completed.  In particular, the concepts of forecast discrimination and  forecast value are discussed based on a simple cost-loss model.  

First, the RUC curve is shown to be the counterpart of the ROC curve when the focus is on a given user rather than on a given event. The areas under the RUC and ROC curves are summary measures of discrimination adapted to quantile and probability forecasts, respectively. Both measures share the same properties, such as non-sensitivity to calibration.

Second, the translation of discrimination ability into value is explored with the help of the value score. The definition of the forecast value is directly adopted from the probability forecast verification framework. Forecast value and forecast potential value are estimated when the decision variable is a quantile forecast, so focusing on a user with a specific cost-loss ratio. The first  is obtained when the forecast is taken at face value and the second  when critical decision criteria are applied.  The value of a quantile forecast can then  be plotted as a function of a range of events of interest, defined for example in terms of base rates. The derived plot is called a quantile value plot and provides a valuable insight into the performance of a quantile forecast. As a real example, the discrimination ability and value of global radiation forecasts from COSMO-DE-EPS are demonstrated over a summer and a winter period. 
 
Finally, it is shown that the overall value of a quantile forecast corresponds to the  quantile skill score with climatology as reference when a constant cost-loss ratio for all outcomes is assumed. In the same spirit as the weighted version of the continuous ranked probability score proposed by \cite{ranjan2011}, a weighted version of the quantile skill score could  be envisaged in order to take into account specific use of quantile forecasts. 

\section*{Acknowledgement}
 This work has been initiated in the framework of the EWeLiNE project (\textit{Erstellung innovativer Wetter- und Leistungsprognosemodelle f\"ur die Netzintegration wetterabh\"angiger Energietr\"ager})  funded by the German Federal Ministry for Economic Affairs and Energy.  

\section*{Appendix}
\label{app:qss}
\subsection*{Overall value and Quantile Skill Score}

From Eqs \eqref{equ:TP} and \eqref{equ:TC}, the difference in expense between climatological and perfect deterministic forecasts can be written as
\begin{equation} 
\begin{split}
T_C - T_P= \dfrac{1}{N} \sum_{i=1}^{N} C   (\Omega_\tau  - \Omega_i) +
 \dfrac{1}{N}  \sum_{i:\Omega_i \geq \Omega_\tau} L   (\Omega_i- \Omega_\tau)
\label{equ:num}
\end{split}
\end{equation}
Considering the relationship  $\tau= 1-\dfrac{C}{L}$ and setting \(L\) equal to 1 in the following demonstration without loss of generality, we obtain
\begin{equation} 
T_C - T_P=   \dfrac{(1-\tau)}{N} \sum_{i=1}^{N} (\Omega_\tau - \Omega_i) + 
 \dfrac{1}{N}  \sum_{i:\Omega_i \geq \Omega_\tau} (\Omega_i-\Omega_\tau)
\end{equation} 
and with some algebra
\begin{equation}
 \begin{split}
T_C - T_P= \dfrac{(1-\tau)}{N} \sum_{i:\Omega_i\leq \Omega_\tau} (\Omega_\tau-\Omega_i)  + 
  \dfrac{\tau}{N} \sum_{i:\Omega_i \geq \Omega_\tau}   (\Omega_i- \Omega_\tau)  
\end{split}
\label{equ:tctp}
\end{equation}  
This  mean expense difference, $T_C - T_P$, corresponds to the definition of the quantile score for a climatological forecast (\(QS_\text{climate}\)).

In the same manner, from Eqs \eqref{equ:TF} and \eqref{equ:TC}, the  difference between climatological forecast expense and the quantile forecast expense is written as
\begin{equation} 
 \begin{split}
T_C - T_F = \dfrac{1}{N} \sum_{i=1}^{N} C   \Omega_\tau +  
\dfrac{1}{N} \sum_{i:\Omega_i \geq \Omega_\tau} L   (\Omega_i -\Omega_\tau)  \\
- \dfrac{1}{N} \sum_{i=1}^N  C   q_{\tau,i} 
- \dfrac{1}{N} \sum_{i:\Omega_i \geq q_{\tau,i}} L  (\Omega_i- q_{\tau,i}) 
\end{split}
\end{equation} 
which becomes after some algebra
\begin{equation} 
\begin{split}
T_C - T_F = 
   \dfrac{(1-\tau)}{N} \sum_{i:\Omega_i\leq \Omega_\tau} (\Omega_\tau  - \Omega_i)  +
   \dfrac{\tau}{N} \sum_{i:\Omega_i \geq \Omega_\tau} (\Omega_i -\Omega_\tau) \\
 -  \bigg( \dfrac{(1-\tau)}{N} \sum_{i:\Omega_i\leq q_{\tau,i}} (q_{\tau,i} - \Omega_i) +
           \dfrac{\tau}{N} \sum_{i:\Omega_i \geq q_{\tau,i}}  (\Omega_i- q_{\tau,i}) \bigg)
\end{split}
\label{equ:tctf}
\end{equation} 
where the first term corresponds to the definition of the quantile score for a climatological forecast (\(QS_\text{climate}\), Eq. \eqref{equ:tctp}), and the second term to the quantile score (\(QS_\text{forecast}\), Eq. \eqref{equ:QSf}). With regard to the definition of the quantile skill score and of the overall value (Eqs \eqref{equ:QSS} and \eqref{equ:Vall}, respectively), we end up with: 
\begin{equation} 
V_{all}=   QSS
\label{equ:vall}
\end{equation}


\end{document}